\documentclass[twocolumn,prd,floatfix,nofootinbib]{revtex4}
\usepackage{mathrsfs}
\usepackage{bbm}
\usepackage{amsmath}
\usepackage{graphicx}
\usepackage{dcolumn}
\usepackage{bm}
\usepackage{amssymb}
\usepackage{latexsym}

\def\be{\begin{equation}}
\def\ee{\end{equation}}
\def\ba{\begin{eqnarray}}
\def\ea{\end{eqnarray}}

\begin{document}

\title{Amplification of Curvature Perturbations in Cyclic Cosmology
}

\author{Jun Zhang\footnote{Email: junzhang34@gmail.com}}
\author{Zhi-Guo Liu\footnote{Email: liuzhiguo08@mails.gucas.ac.cn}}
\author{Yun-Song Piao\footnote{Email: yspiao@gucas.ac.cn}}

\affiliation{College of Physical Sciences, Graduate School of
Chinese Academy of Sciences, Beijing 100049, China}

\begin{abstract}

We analytically and numerically show that through the cycles with
nonsingular bounce the amplitude of curvature perturbation on
large scale will be amplified and the power spectrum will be
redden. In some sense, this amplification will eventually destroy
the homogeneity of background, which will lead to the ultimate end
of cycles of global universe. We argue that for the model with
increasing cycles, it might be possible that a fissiparous
multiverse will emerge after one or several cycles, in which the
cycles will continue only at corresponding local regions.

\end{abstract}

\maketitle

Recently, a cosmological cyclic scenario, in which the universe
experiences the periodic sequence of contractions and expansions
\cite{Tolman}, has been rewaked \cite{STS}, and brought the distinct
insights into the origin of observable universe. There have been
lots of studies on cyclic or oscillating universe models
\cite{BD},\cite{KSS},\cite{Piao04},\cite{Lidsey04},\cite{CB},\cite{Xiong},\cite{Xin},\cite{LS},
\cite{Biswas},\cite{Biswas1},\cite{Cai0906}, also \cite{NB} for a
review. In general, it is thought that the background of cyclic
universe is homogeneous cycle by cycle all along. However, it has
been noticed that in the contracting phase the amplitude of
curvature perturbation on super Hubble scale is increased, while is
nearly constant in the expanding phase. Thus the net result is that
the amplitude of perturbation is amplified, which seems to go along
cycle by cycle \cite{Piao0901}.

Whether this amplification of perturbation actually occurs is
interesting, since it will lead the global configuration of cyclic
universe unexpected colorful \cite{Piao1001}. In this paper, we will
analytically and numerically show the change of power spectrum of
curvature perturbation through one or several cycles. We will
simulate the effect of the increasing of its amplitude on the global
configuration of cyclic universe, which might dramatically alter our
conventional perspective for cyclic universe.

\begin{figure}
\includegraphics[width=8.0cm]{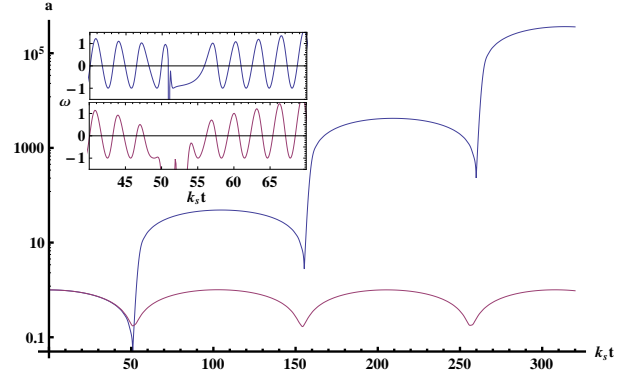}
\caption{\label{fig:a-w} The evolution of scale factor in cyclic
universe models. The blue line denotes that with increasing cycles.
The red line denotes that with equal cycles. The time is scaled by
$k_s$ which is an adjustable parameter to make the models be able to
be consistent with the observable universe.}
\end{figure}

We begin with the cyclic universe models in Fig.\ref{fig:a-w}, in
which the nonsingular bounce is implemented by introducing a field
with negative energy, which is minimally coupled to gravity.
During the expansion and contraction, the universe is dominated by
a normal scalar field, which oscillates in a quadratic potential
and offers matter-dominated background. With the shrinking of the
scale factor, the ghost field becomes dominated and finally leads
the bounce take place. In the inset of Fig.1, the corresponding
state parameters $\omega \equiv p/\rho$ of models are plotted. It
can be found that, due to the oscillation of the dominated scalar
field, $\omega$ is oscillating with $\langle\omega\rangle \simeq
0$ \cite{Turner}, which is matter-liked, during the contraction
and expansion. While near the bounce, $\omega$ crosses $-1$ in a
very short time.

In general, the introduction of field with negative energy is only
the approximative simulation of a fundamental theory below certain
physical cutoff. Pointed by \cite{cline}, for a ghost with a
minimal coupling to gravity, as same case in our paper, the cutoff
$\Lambda$ is constrained by observations of the diffuse gamma ray
background with $\Lambda < 3 MeV$. Therefore, we can estimate the
validity of our model by consider the valid physical momentum
space of perturbation. Given by the cutoff $\Lambda$, the upper
limit of the comoving wave numbers is $k_c \sim a\Lambda/k_s$
where $k_c$ is in units of $k_s$, an adjustable parameter that can
be determined by fit the time of expansion, which is $40/k_s$, see
Fig.1, to the age of observable universe, $40/k_s \sim 1/H_0$.
When we use the present Hubble rate, $H_0 \sim 10^{-33}eV$, we can
find that, during the bouncing, the upper limit of the
perturbations' comoving wave numbers is $k_c \sim 10^{26}$ which
is much larger than the largest wave number, $10^6$, which is
chose in following figures', discussed in this paper. In this
sense, the validity of the background model used in this paper is
well guaranteed. However, one should remember that the appearance
of phantom field is only an artificial approximation. Recently,
the nonsingular bounce has been obtained in nonlocal higher
derivative theories of gravity \cite{BMS},\cite{BKM}, which can be
ghostfree.

The primordial perturbation generated during the contraction with
$w\simeq 0$ is scale invariant \cite{Wands99},\cite{FB},\cite{S}.
There also are some studies how the perturbations go through such a
nonsingular bounce \cite{AW},\cite{BV},\cite{FPP},\cite{Cai0810}.
The cycle plotted as the blue line is increased, because there is an
inflationary epoch after the bounce in each cycle, the corresponding
observable signals have been studied in e.g.
\cite{Piao0308},\cite{Piao0310},\cite{Piao0501}. The models with
increasing cycle can be also obtained in e.g.
\cite{BD},\cite{KSS},\cite{Biswas},\cite{Biswas1}, in which the
entropy is increased. However, the results of change of power
spectrum of curvature perturbation through cycles are not altered
qualitatively by how the increasing cycle is implemented. The
examples in Fig.1 will only serve the purpose of numerical
simulations in the following.

We will regard the turnaround time as the beginning of a cycle,
which can be denotes as $t_T^j$ of the $j^{th}$ cycle. In each
cycle the universe will orderly experience the contraction,
bounce, and expansion, and then arrive at the turnaround, which
signals the end of a cycle. The beginning and the end of each
phase of the $j^{th}$ cycle can be denoted as $t_{Ci}^j$,
$t_{Ce}^j$, $t_{Ei}^j$, $t_{Ee}^j$, and the bounce is 
$t_B^j$.

The evolution of curvature perturbation under this cyclic
background can be simply showed as follows. The motive equation of
the curvature perturbation $\zeta$ in the momentum space is \be
u_k^{\prime\prime} +\left(k^2-{z^{\prime\prime}\over z}\right) u_k
= 0 ,\label{uk}\ee where $u_k \equiv z\zeta_k$
\cite{Muk},\cite{KS}, and \cite{Mukhanov} for details, the prime
denotes the derivative for the conformal time $\eta$ and
$z={a\over H}({|{\dot H}|\over 4\pi G})^{1/2}$. When $k^2 \gg
z^{\prime\prime}/z$, i.e. the perturbations are deep inside the
Hubble radius, it is obviously that $u_k$ will oscillate with a
constant amplitude. When $k^2\ll z^{\prime\prime}/z$, i.e. the
perturbations are on super Hubble scale, Eq.(\ref{uk}) has general
solution $u_k$, which gives $\zeta_k \simeq C_1+C_2\int{d\eta\over
z^2}$, e.g.\cite{Mukhanov}, where $C_1$ and $C_2$ are constant for
fixed $k$. In general, for the contraction phase $a\sim
(t_B-t)^{n}$, $a\sim (\eta_B-\eta)^{n\over 1-n}$ can be obtained.
Thus during the contraction with $n>{1\over 3}$, as $t \rightarrow
t_B$, the amplitude of $\zeta$ will be dominated by $C_2$ term,
\begin{eqnarray} \zeta_k \sim \int{d\eta\over z^2}\sim
\left(t_B-t\right)^{1-3n}, \label{zeta}\end{eqnarray} since
$\int{d\eta\over z^2}\simeq \int{d\eta\over a^2}\sim
(\eta_B-\eta)^{1-3n\over 1-n}$, which is increased on large scale.

While for the expanding phase $a\sim t^n$, $a\sim
(\pm\eta)^{n\over 1-n}$ can be obtained, where the plus is for
${1\over 3}<n<1$ and the minus for $n>1$. In these both cases the
$C_2$ term is decreasing. Thus the $C_1$ term dominates $\zeta_k$,
which is constant for fixed $k$. This means that for a cycle of cyclic universe
during the contraction $\zeta_k$ is increased on super horizon scale, up
to the end of contracting phase in corresponding cycle, while
during the expansion it becomes constant. Thus the net result is
that $\zeta_k$ on large scale is amplified, which is inevitable.

\begin{figure}
\includegraphics[width=8.5cm]{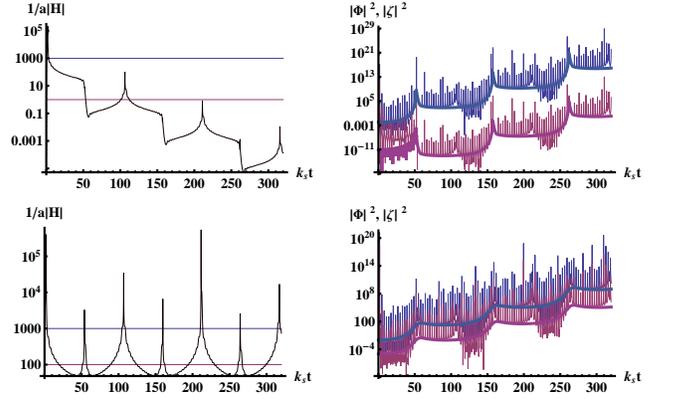}
\caption{\label{fig:k} The evolutions of perturbations through
cycles given in Fig.1. The left upper and lower panels correspond
to the models with increasing cycle and equal cycle in Fig.1,
respectively. The black line denotes the evolutions of the Hubble
radius ${1\over a|H|}$, and the blue and red lines denote the
evolutions of the perturbations with the modes $10^{-3}k_s$ and
$k_s$, respectively, which crosse the Hubble radius at different
time. The evolutions of corresponding modes are given in the right
upper and lower panels, respectively.
The evolutions of $|\Phi_k|^2$ are denoted with thick lines, while
that of $|\zeta_k|^2$ are denoted with thin lines. We can see that
$\zeta_k$ is increased cycle by cycle. }
\end{figure}

The spectrum index of $\zeta$ obtained by Eq.(\ref{uk}) is
$n_\zeta-1=3-\left|{3n-1\over n-1}\right|$, e.g.\cite{Piao0404},
which is scale invariant for $n\gg 1$, i.e. inflation, and for
$n\simeq {2\over 3}$, i.e. the contraction with $w\simeq 0$
\cite{Wands99},\cite{FB},\cite{S}.
During the contraction, when $n>{1\over 3}$, the $C_2$ term
dominates,
while when $0<n <{1\over 3}$, $\zeta$ is not dominated by the
$C_2$ term, i.e. its increasing mode, see Eq.(\ref{zeta}). Thus in
this case, it seems that the curvature perturbation is not
amplified. $n\simeq 0_+$ corresponds to that in \cite{STS},
however, see \cite{KS1} for the case with changed $w$.

The net amplification for the perturbation modes, which are on
super horizon scale in the $j^{th}$ and $j^{th}+1$ cycles all
along, can be estimated as follows. In the $j^{th}$ cycle, after
the bounce, $\zeta$ is given by Eq.(\ref{zeta}), which will be
unchanged up to the end of the $j$ cycle. Then the universe enters
into the $j^{th}+1$ cycle, during the contraction of the
$j^{th}+1$ cycle, $\zeta$ will continue to increase. Thus at
certain time during the contraction, we have \ba
\zeta^{j+1}(t^{j+1})&\simeq &
\left(\frac{t_{B}^{j+1}-t^{j+1}}{t_{B}^{j+1}-t_{Ci}^{j+1}}\right)^{1-3n_{j+1}}\zeta^{j}(t_{Ce})\nonumber\\
&\sim &
\left(\frac{t_{B}^{j+1}-t^{j+1}}{t_{B}^{j+1}-t_{Ci}^{j+1}}\right)^{1-3n_{j+1}}
\left(t_{B}^j-t_{Ce}^j\right)^{1-3n_{j}}
\nonumber\\
&\sim &  \exp{\left({(3n^{j+1}-1){\cal N}^{j+1}\over
1-n^{j+1}}\right)}H_{Ce}^j
, \label{zeta1}\ea since $H\sim 1/(t_B-t)$, which means that for
these modes still staying on super horizon scale the amplitude of
the spectrum will be amplified with same rate after each cycle. In
the third line \be {\cal N}^{j+1}=\ln\left({a^{j+1}H^{j+1}\over
a_{Ci}^{j+1}H_{Ci}^{j+1}}\right)\simeq
(1-n^{j+1})\ln\left({H^{j+1}\over H_{Ci}^{j+1}}\right) \ee is the
e-folding number for the primordial perturbation generated during
the contraction of the $j^{th}+1$ cycle. When $w\simeq 0$, i.e.
$n^{j+1}\simeq {2\over 3}$, Eq.(\ref{zeta}) becomes
$\zeta^{j+1}(t^{j+1})\simeq e^{3{\cal N}^{j+1}}H_{Ce}^j$, which is
consistent with that in \cite{Piao0901,Piao1001}.

We numerically show the evolutions of the perturbation modes
$\zeta_k$ in details under the backgrounds in Fig.\ref{fig:a-w},
which are plotted in proper time. This actually can be manipulated
by numerically solving the equation of the metric perturbation
$\Phi_k$ with $c_s^2=1$. Then the evolution of $\zeta_k$ is obtained
by using $\zeta={2\over 3}({H^{-1}{\dot \Phi}+\Phi\over 1+w})+\Phi$.
We can clearly see that $\zeta_k$ is increasing during the
contraction and is constant during the expansion, which thus is
amplified cycle by cycle. The results are consistent with above
discussions, and also the numerical results in
\cite{AW},\cite{Cai0810} for one cycle.

We neglected the effect of entropy perturbation in this simulation.
However, the inclusion of the entropy perturbation will not change
the result essentially, since it is generally not important. Though
in the background evolution of Fig.1, the nonsingular bounce is
implemented by introducing the field with negative energy, the
numerical results obtained might be applicable for other cyclic
models with nonsingular bounce. The differences are that in these
models around the bounce $c_s^2$ might change with time and the
higher order terms of $\Phi$ appear. However, if the correction for
the equation of $\Phi$ is important only around the bounce, the
results are not expected to be altered by these difference.

\begin{figure}
\includegraphics[width=8.5cm]{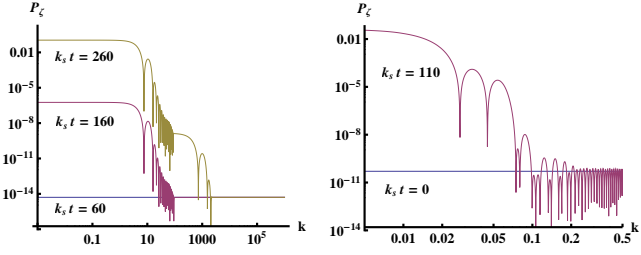}
\caption{\label{fig:ns} The left panel is the power spectrum of
perturbations for model with increasing cycle in Fig.1 at
different time. The times chose to plot the power spectrum have
been signaled in the figure, $k_st=60$ is that after the bounce in
the $j^{th}$ cycle and $k_st=160$ is that after the bounce in the
$j^{th}+1$ cycle and $k$ is in the units of $k_s$ as in Fig.1. The
right panel is that of perturbations for model with equal cycle in
Fig.1. We can see that after one cycle the spectrum will be redden
on corresponding scale.}
\end{figure}

The power spectrum of $\zeta$ is ${\cal P}_{\zeta}(k)
=\frac{k^3}{2\pi^2}|\zeta_k|^2$. We assume that the power spectrum
of $\zeta$ after the bounce of the $j^{th}$ cycle is ${\cal
P}_{\zeta}(k,t_B^j)$. Thus in term of Eq.(\ref{zeta}), at certain
time $t^{j+1}$ during the contraction of the $j^{th}+1$ cycle, the
power spectrum of the perturbation modes, which are all along on
super horizon scale, is given by
\begin{eqnarray}
\label{P1} {\cal P}_{\zeta}^{j+1}(k,t^{j+1})
\simeq&&\left(\frac{t_{B}^{j+1}-t^{j+1}}{t_{B}^{j+1}-t_{Ci}^{j+1}}\right)^{2-6n_{j+1}}
{\cal P}_{\zeta}^{j}(k,t_B^j),
\end{eqnarray}
which means that the amplitudes of these modes are amplified, but
the spectrum shape is unchanged. There are also the perturbation
modes, which will enter into the horizon during the $j^{th}$ cycle and
then leave it during the contraction of the $j^{th}+1$ cycle, see the red lines in
right panels of Fig.2.
Inside the Hubble radius, $u_k$ is constant. Thus we have
$\zeta_k=u_k/z\sim u_k/a$. This means that the amplitude of
$\zeta_k$ is decreasing during the expansion and increasing during
the contraction on subhorizon scale. The power spectrum of these modes can be
estimated as \ba \label{P2} {\cal P}_{\zeta}^{j+1}(k,t^{j+1})
&\simeq &
\frac{(t_{B}^{j+1}-t_{Ci}^{j+1})^{2n^{j+1}}}{(t_{Ee}^j-t_{B}^j)^{2n^{j}}}\left({t_{B}^{j+1}-t^{j+1}}\right)^{2-6n^{j+1}}
\nonumber\\ & & \left({kk_B^{j+1}\over k_B^{j+1}-k}
\right)^{\frac{4n^{j+1}-2}{n^{j+1}-1}}\left({kk_B^{j}\over
k_B^{j}-k} \right)^{\frac{2n^{j}}{n^{j}-1}}{\cal
P}_{\zeta}^{j}(k,t_B^{j}), \nonumber\\ &\sim & k^{
\Delta n_s}{\cal P}_{\zeta}^{j}(k,t_B^{j}), \ea for
$t_{Ci}^{j+1}<t^{j+1}<t_{Ce}^{j+1}$, where $k_B^j$, $k_B^{j+1}$ are
the modes crossing the horizon at the bounce time $t_B^j$. In
general, the change of spectrum is not exactly power law. The second
line is obtained only when $k\ll k_B^{j}, k_B^{j+1}$. We can see
that for these perturbations modes, not only the amplitude of the
spectrum is increasing after each cycle, but also the shape of the
spectrum is changed with
\begin{eqnarray}
\Delta n_\zeta \simeq
{\frac{4n^{j+1}_C-2}{n^{j+1}_C-1}}+{\frac{2n^{j}_E}{n^{j}_E-1}},
\end{eqnarray}
where $n^j_E$ and $n^{j+1}_C$ denote that in the expanding phase
of the $j^{th}$ cycle and that in the contracting phase of the
$j^{th}+1$ cycle, respectively, which are generally not equal. In
general, for the interesting value of $n^j_E$ and ${1\over
3}<n^{j+1}_C<1$, $\Delta n_\zeta$ is negative, thus the spectrum
will redshift. However, it is possible that $\Delta n_\zeta\simeq
0$ by choosing special $n^j_E$ and $n^{j+1}_C$, which might has
interesting application, e.g.\cite{BMS1}. In this calculation, we
neglected the effect of the transfer function, which is dependent
on the matter content in corresponding cycle. However, the
evolutive behaviors of spectrum obtained here are not altered
qualitatively. When $n^j_E=n^{j+1}_C=n$, the shape of spectrum is
changed as $\Delta n_\zeta \simeq {\frac{6n-2}{n-1}}$, which is
consistent with the result in \cite{B0905}.

We numerically show the change of the power spectrum through one
cycle in Fig.\ref{fig:ns} under the backgrounds in
Fig.\ref{fig:a-w}. The power spectrum ${\cal
P}_{\zeta}^{j}(k,t_B^j)$ of $\zeta$ after the bounce in the $j$
cycle is scale invariant, since $n\simeq {2\over 3}$. During the
contraction of the $j^{th}+1$ cycle, for the model with equal cycle
the spectrum will redshift. While for that with increased cycle, the
shape of spectrum on large scale is unchanged, i.e. still scale
invariant, only its amplitude is amplified, since these modes are on
super horizon scale during the contraction of the $j^{th}$ cycle and
$j^{th}+1$ cycle all along, however, the spectrum of the modes on
middle scale will redshift, and the rate of tilt can be estimated as
$\Delta n_\zeta\simeq 6$. The spectrum on small scale is scale
invariant, because these modes are newly generated in the $j^{th}+1$
cycle. These results are consistent with Eqs.(\ref{P1}) and
(\ref{P2}), which obviously reflect the changes of power spectrum
through cycles.

Therefore, for ${1\over 3}<n< 1$ the amplitude ${\cal
P}_{\zeta}^{1/2}$ of perturbation on large scale will inevitably
arrive at ${\cal P}_{\zeta}^{1/2}\sim 1$ at certain time of cycles.
We only show it $\sim 0.1$ in Fig.\ref{fig:ns}, since the linear
perturbation approximation is not reliable when ${\cal
P}_{\zeta}^{1/2}\sim 1$, and in this case, the nonlinear effect will
become important and the coupling between modes has to be
considered. However, it can be expected that the enhancement of
nonlinear effect will bring ${\cal P}_{\zeta}^{1/2}$ to 1 faster.
This means that on the corresponding scale the perturbation will be
large enough so that will destroy the homogeneity of background. We
illustrate this effect of the perturbation on background by
transforming ${\cal P}_{\zeta}^{1/2}$ in Fig.3 into $\zeta(\vec{x})$
in position space plotted in Fig.4.

We can see that when ${\cal P}_{\zeta}^{1/2}\rightarrow 1$ the
universe is fragmentized into large number of small regions. In
this case, the initially global homogeneous universe will become
highly inhomogeneous. Thus it can be hardly imagined that the
different regions of global universe will evolve synchronously,
even if it is homogenous in the previous cycle. Thus the cycle of
global universe will ultimately end. The similar phenomenon was
also discussed in \cite{Erick}, however, in which since the bounce
is singular the special mode mixing has to be applied for the
amplification of $\zeta$, e.g.\cite{DH},\cite{DV},\cite{TBF}.
Here, the bounce builded is nonsingular, thus the evolution of
perturbations through cycles can be exactly studied by numerical
method.

In general, as long as the
curvature perturbation on large scale is amplified cycle by cycle actually occurs,
it is hardly possible that the global universe is eternal cyclic.
We only consider the evolution of curvature perturbation, the
effect of the increasing of other perturbations on cyclic universe has been also
studied, e.g.\cite{BY}. These results mean that the reliability of
some models of cyclic universe might need to be reevaluated.

\begin{figure}
\includegraphics[width=8.5cm]{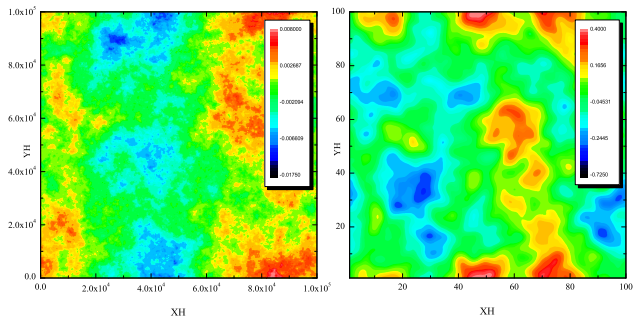}
\caption{\label{fig:4} $\zeta(\vec{x})$ in position space, which
reflects the inhomogeneity of background. The left panel is for
the model with increasing cycles at the time $k_st=260$, while the
right panel is for that with equal cycles at the time $k_st=110$,
and the length scale is in unit of the Hubble radius at that
time.}
\end{figure}

However, it can be noticed that for the universe with increasing
cycle, the regions split can be larger. When the length of local
regions split is larger than the Hubble scale at the corresponding
time, these regions will possibly evolve independently, as long as
inside the corresponding regions the background is homogeneous. In
principle, such different local regions correspond to different
universes \cite{CH},\cite{WMLL}, each of which is controlled by
local physical equations and might be fragmentized again after
itself cycles. In this case, the number of local universes will
increase cycle by cycle. Thus we can have a cyclic multiverse
scenario, as argued in \cite{Piao0901},\cite{Piao1001}. This
argument in some sense indicates that in cyclic cosmology a
gradually increasing cycle is significant for the continuance of
cycle, however, in this case, the cycle will continue only at local
regions, which is homogeneous. This scenario can be distinguished
from that in chaotic eternal inflation \cite{83,86}, in which the
inflationary multiverse is induced by the large quantum fluctuation
of inflaton field, which occurs efold by efold. However, the cyclic
multiverse is induced by the cyclic amplification of perturbation on
large scale, which is in classical sense, and occurs cycle by cycle.

In conclusion, we have analytically and numerically showed that
through the cycles with nonsingular bounce the amplitude of
curvature perturbations on large scale will be amplified and the
power spectrum will be redden. In some sense, this amplification
will eventually destroy the homogeneity of background, which might
lead to the ultimate end of cycles of global universe. However, it
can be argued that for the model with increasing cycle, the global
universe will possibly evolve into a fissiparous multiverses after
one or several cycles, in which the cycles will continue only at
corresponding local regions, inside which the background is
homogeneous.

\textbf{Acknowledgments} We thank for Y.F. Cai and A. Mazumdar
helpful discussions and comments. This work is supported in part
by NSFC under Grant No:10775180, 11075205, in part by the
Scientific Research Fund of GUCAS(NO:055101BM03), in part by
National Basic Research Program of China, No:2010CB832804


\end{document}